\documentstyle[12pt]{article}

\def\Msun{\ifmmode{~M_\odot}\else$M_\odot$~\fi}
\def\kms{\ifmmode{$~km\thinspace s$^{-1}}\else km\thinspace s$^{-1}$\fi}
\def\etal{\ifmmode{${\it et al.}$}\else {\it et al.}\fi}

\def\ga{\mathrel{\mathchoice {\vcenter{\offinterlineskip\halign{\hfil
$\displaystyle##$\hfil\cr>\cr\noalign{\vskip1.5pt}\sim\cr}}}
{\vcenter{\offinterlineskip\halign{\hfil$\textstyle##$\hfil\cr>\cr
\noalign{\vskip1.0pt}\sim\cr}}}
{\vcenter{\offinterlineskip\halign{\hfil$\scriptstyle##$\hfil\cr>\cr
\noalign{\vskip0.5pt}\sim\cr}}}
{\vcenter{\offinterlineskip\halign{\hfil$\scriptscriptstyle##$\hfil
\cr>\cr\noalign{\vskip0.5pt}\sim\cr}}}}}

\begin{document}
\baselineskip=24pt

\title{Kinematics of the Outer Stellar Halo}

\author{Chris Flynn{$^{1,2}$}, J.~Sommer-Larsen{$^3$ \&
P.~R.~Christensen{$^{3,4}$}}\\ \\ \\ 
{$^1$}NORDITA, \\Blegdamsvej 17, DK-2100 Copenhagen {\O}, Denmark\\\\
{$^2$}Tuorla Observatory, \\V\"ais\"al\"antie 20, FIN-21500 Piikki\"o,
Finland \\\\
{$^3$}Theoretical Astrophysics Center,\\Blegdamsvej 17, DK-2100
Copenhagen {\O}, Denmark \\\\
{$^4$}The Niels Bohr Institute,\\Blegdamsvej 17, DK-2100 Copenhagen
{\O}, Denmark}

\date{ }
\maketitle

\section*{ }

\section*{Summary}
\baselineskip=12pt

  We have tested whether the simple model for the kinematics of the
Galactic stellar halo (in particular the outer halo) proposed by
Sommer-Larsen, Flynn and Christensen (SLFC) is physically realizable,
by directly integrating particles in a 3-D model of the Galactic
potential. We are able to show that the SLFC solution can be realized
in terms of a distribution of particles with stationary statistical
properties in phase-space.  Hence, the SLFC model, which shows a
notable change in the anisotropy from markedly radial at the sun to
markedly tangential beyond about Galactocentric radius $r=20$ kpc,
seems a tenable description of outer halo kinematics.

\newpage
\baselineskip=12pt

\section{Introduction}

  In recent years various techniques have been developed to isolate
field halo stars in situ, so that increasingly, samples of these stars
are becoming available with which to address questions about the
kinematic structure of the halo. Radial velocity and distance
measurements of these objects provide important constraints on models
of the kinematics and more indirectly the distribution of the dark
matter in which the visible galaxy is apparently embedded.

  Flynn, Sommer-Larsen and Christensen (1994) have developed
techniques using broadband photoelectric photometry and medium
dispersion spectroscopy for isolating Blue Horizontal Branch (BHB)
stars in the outer Galactic halo and have assembled a catalog in four
fields of about 100 stars with line-of-sight velocities and distances
(Flynn et. al. 1995).  Sommer-Larsen, Flynn and Christensen (1994,
hereafter SLFC) developed a model of the kinematics of the outer halo,
which fits the observations surprisingly well, given the simple nature
of the model.  They assumed that the outer stellar halo is a round,
non-rotating system with a density falloff with Galactocentric radius
like $r^{-3.4}$ and that the Galactic rotation curve is flat to large
Galactic radii ($r >> R_\odot $) where $R_\odot$ is the sun's distance
from the Galactic center.  The Jeans equation was then solved for the
radial and one-dimensional tangential velocity dispersions ($\sigma_r,
\sigma_t$) as functions of $r$ by fitting the observed line-of-sight
velocity dispersions as a function of line-of-sight distance.

  Their results indicate that the halo kinematics in the outer halo
are {\it tangentially} anisotropic, whereas they are {\it radially}
anisotropic near the sun.  Specifically, near the sun, $\sigma_r
\approx 140$ \kms~and $\sigma_t \approx$ 90-100 \kms, whereas SLFC
find a major kinematic change in the outer halo $(r \ga 10-20)$ kpc,
where $\sigma_r \approx 80-100$ \kms~and $\sigma_t \approx $130-150
\kms.

  The solution to the Jeans equation only provides the second moments
of the velocity distribution in the halo. To demonstrate that the
solution for $(\sigma_r, \sigma_t)$ is actually physical, one must
show that the velocity dispersions can be realized in terms of a
stationary phase-space distribution function $f$ which is everywhere
non-negative in phase-space.  Although there are methods in the literature for
recovering the distribution function from simple potential-density
pairs and the velocity dispersions (see Binney and Tremaine (1987)
pp255 and references therein), the extension of these methods to the
case of the Galactic potential is a formidable analytic task.  Since
we would nevertheless like to test that the SLFC model is physical, we
have taken in this paper a less ambitious and more direct approach: we
place a large number of test particles with the kinematic
characteristics of the SLFC type model (using a simple assumption
about their velocity distribution) into a realistic 3-dimensional
model of the Galaxy's potential, and integrate the orbits over a
Hubble time, allowing the system to phase mix.  We show in this way
that a stationary system can be realized with the kinematic and
spatial properties of the SLFC model.

  In section 2 we develop a realistic model of the Galactic potential.
In section 3 we describe our simulations of the SLFC kinematics, and
we discuss the results and draw conclusions in section 4.

\section{A Galactic Potential}

  We have developed a multi-component model of the Galactic potential
which we have matched with good accuracy to Galactic parameters, such
as the rotation curve, local disk density and disk scale length.

  The Galactic potential $\Phi(R,z)$ is modeled in cylindrical
coordinates where $R$ is the planar Galactocentric-radius and $z$ is
the distance above the plane of the disk. The total potential $\Phi$
is modeled by the sum of the potentials of the dark halo $\Phi_H$, a
central component $\Phi_C$ and disk $\Phi_D$:

$$\Phi = \Phi_H + \Phi_C + \Phi_D.$$ 

The potential of the dark halo $\Phi_H$ is assumed to be spherical and
of the form

$$\Phi_H = {1\over{2}}V_H^2 {\rm ln}(r^2 + r_0^2)$$

\noindent where $r$ is the Galactocentric radius ($r^2 = R^2 +
z^2$). The potential has a core radius $r_0$ and $V_H$ is the circular
velocity at large $r$ (i.e. relative to the core radius).

The potential of the central component $\Phi_C$ is modeled by two
spherical components, representing the bulge/stellar-halo and an inner
core component:

$$\Phi_C = - {{G M_{C_1}}\over{\sqrt{r^2+r^2_{C_1}}}} - {{G
M_{C_2}}\over{\sqrt{r^2+r^2_{C_2}}}}.$$

Here $G$ is the gravitational constant, $M_{C_1}$ and $R_{C_1}$ are
the mass and core radius of the bulge/stellar-halo term and $M_{C_2}$ and
$R_{C_2}$ are the mass and core radius of the inner core.

The disk potential $\Phi_D$ is modeled using an analytical form which
is a combination of three Miyamoto-Nagai potentials (Miyamoto and
Nagai 1975)

$$\Phi_D = \Phi_{D_1} + \Phi_{D_2} + \Phi_{D_3}.$$

where

$$\Phi_{D_n} = {-G M_{D_n} \over{
\sqrt{(R^2+(a_n+\sqrt{(z^2+b^2)})^2)}}} ~~~~~~~~~~ n = 1,2,3 $$

Here the parameter $b$ is related to the disk scale height, $a_n$ to
the disk scale length and $M_{D_n}$ are the masses of the three disk
components.  Adopting a single Miyamoto-Nagai potential for the disk
and reasonable dark halo potentials leads to the problem that the disk
scale length is too short (typically by a factor of 2) for well
fitting rotation curves.  We have circumvented this difficulty by
combining three Miyamoto-Nagai disks of differing scale lengths and
masses (the parameters for which are shown in Table 1). The parameters
have essentially been chosen to be consistent with the dynamical mass
measurements of the disk surface density at the solar circle, while
maintaining a good fit to the rotation curve and a realistic disk
scale length. Note particularly that for the $n=2$ component the mass
$M_2$ is negative.  The densities resulting from the disk potential
alone, as well as the total potential, are nevertheless positive
everywhere.

\smallskip
  \begin{table} 
    \begin{center} 
      \begin{tabular}{l|rr} \noalign{\hrule}
  Component          & Parameter  & Value \\ \noalign{\hrule}
  Dark Halo          & $r_0$      & 8.5 kpc   \\
                     & $V_H $     & 220 \kms \\ \noalign{\hrule}
  Bulge/Stellar-halo & $r_{C_1}$  & $2.7$ kpc   \\
                     & $M_{C_1}$  & $3.0\times 10^9$ \Msun \\
  Central comp.      & $r_{C_2}$  & $0.42$ kpc   \\
                     & $M_{C_2}$  & $1.6\times 10^{10}$ \Msun\\\noalign{\hrule}
  Disk               & $b$        & $0.3$ kpc   \\
                     & $M_{D_1}$  & $6.6\times 10^{10}$ \Msun\\
                     & $a_1$      & $ 5.81$ kpc \\
                     & $M_{D_2}$  & $-2.9\times 10^{10}$ \Msun\\
                     & $a_2$      & $17.43$ kpc \\
                     & $M_{D_3}$  & $3.3\times 10^{9}$ \Msun\\
                     & $a_3$      & $34.86$ kpc \\\noalign{\hrule}
      \end{tabular}
    \end{center}
  \caption{Adopted Parameters for the Galactic Potential}
  \end{table}
\smallskip
 
  The rotation curve of our model is shown in Figure 1, together with the
contributions from the various components. Assuming $R_\odot = $ 8
kpc, the local disk surface density of matter $\Sigma_{\rm
Disk}(R=R_\odot)$ is 51 \Msun pc$^{-2}$, consistent with recent
dynamical measurements which measure about 50 \Msun pc$^{-2}$ (Kuijken
and Gilmore 1989, Flynn and Fuchs 1994). The local midplane density of
the disk (i.e at $R=R_\odot$) is 0.09 \Msun pc$^{-3}$.  The surface
density $\Sigma_{\rm Disk}$ is plotted as a function of $R$ in Figure
2, and the disk is seen to be exponential over a large range in $R$;
it has a scale length of 4.1 kpc, in good agreement with observation
(see e.g. Lewis and Freeman 1989 and references therein).  The disk
has a uniform scale height with $R$ of about 260 pc.

\begin{figure}
\input epsf
\centering
\leavevmode
\epsfxsize=0.65
\columnwidth
\epsfbox{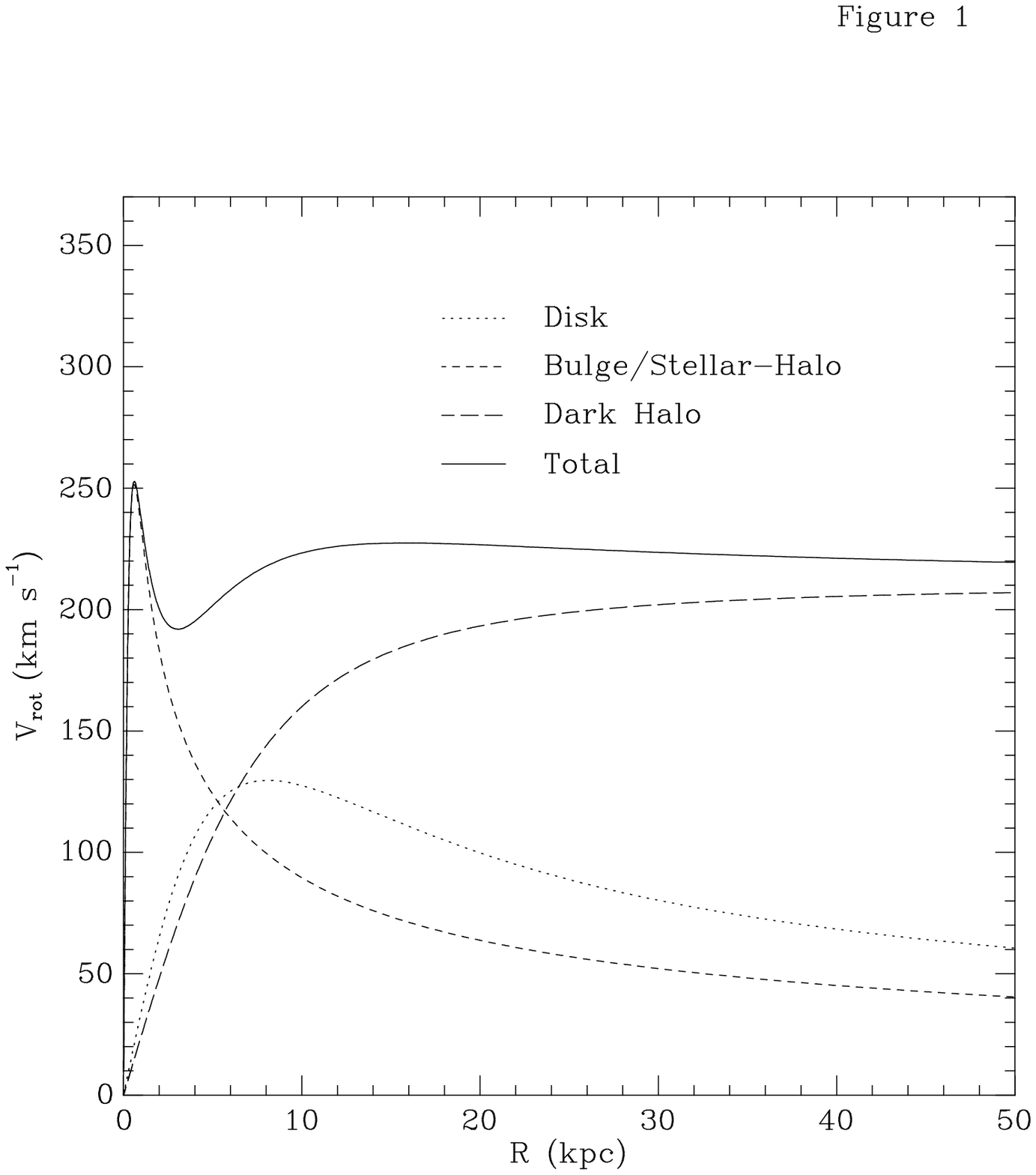}
\caption{The rotation curve of our model galaxy. The
solid line is the total rotation curve, the other lines indicate the
partial contributions to the rotation curve by the three components.}
\end{figure}

\begin{figure}
\input epsf
\centering
\leavevmode
\epsfxsize=0.65
\columnwidth
\epsfbox{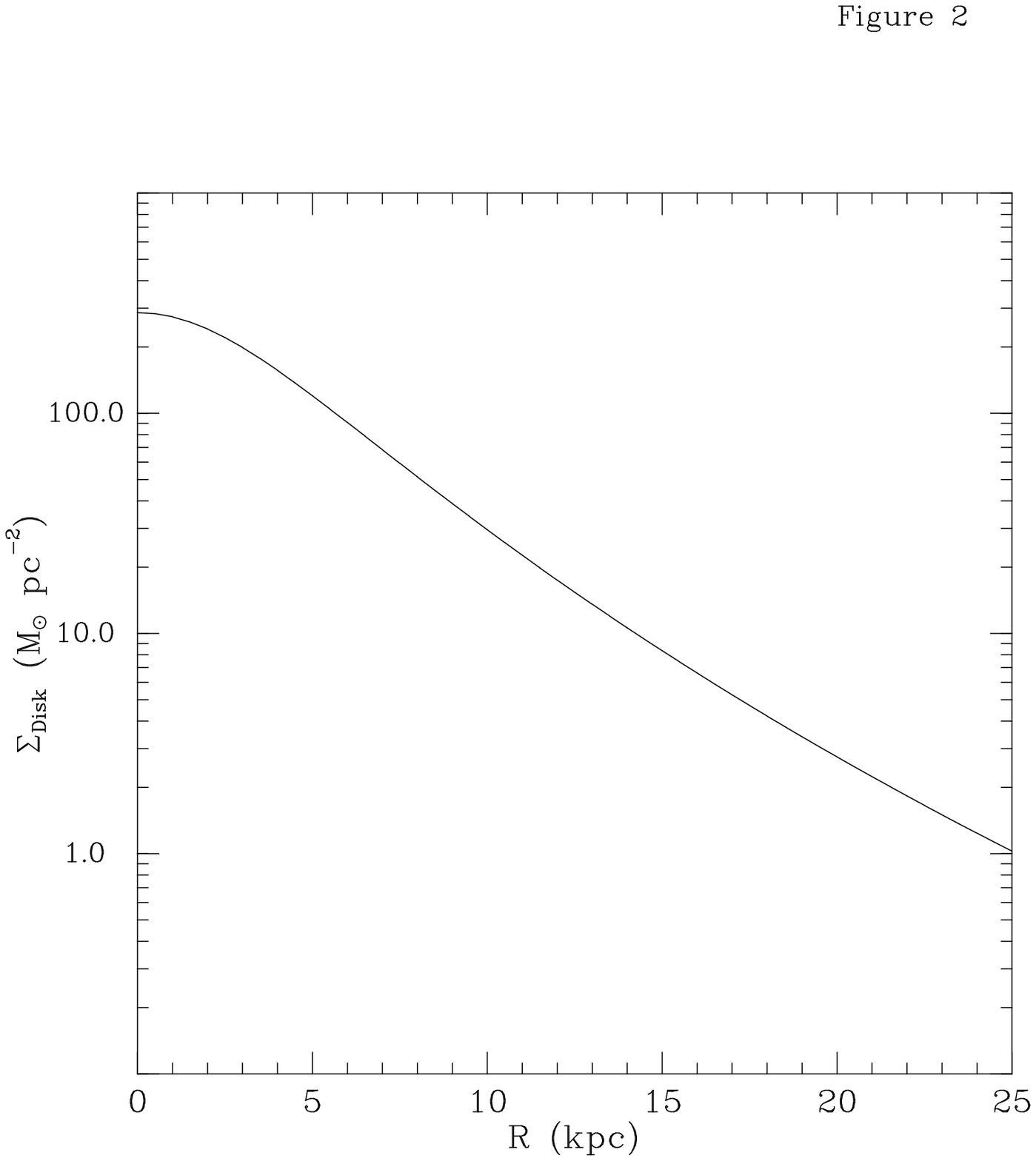}
\caption{Surface density $\Sigma_{\rm Disk}$ in
\Msun pc$^{-2}$ of the disk component of the model potential as a
function of radius in the plane, $R$. This density falloff is close to
an exponential with scale length of 4.1 kpc.}
\end{figure}

\section{Numerical Methods}

  SLFC obtained a best fit solution for the second moments of the
radial and tangential velocity distributions of the stellar halo for
$r \ge R_\odot$. We now describe our simulations in the potential
described above, and show there is a strong indication that the SLFC
model is physically realizable.

  We started by testing whether a phase-space distribution function
with a Gaussian velocity distribution and second velocity moments
given by the SLFC model is stationary in our model potential, i.e. the
phase-space distribution function does not depend explicitly on time.
We set up $5\times 10^4$ tracer particles in the potential described
above, with a density distribution falling like $r^{-3.4}$, truncated
inside 0.1 kpc and outside 200 kpc. Initial particle velocities were
randomly drawn from Gaussian distributions with radial and tangential
velocity dispersions $\sigma_r(r)$ and $\sigma_t(r)$ as specified by
the SLFC (from their equations 3 and 4; see also their Figure 2).  The
orbits of the particles were integrated using a Runge-Kutta scheme
with an adaptive step size (Press et. al. 1992) over a period of 15
Gyr, and the density and velocity dispersions of the particles as
functions of $r$ were determined at 1 Gyr intervals.  The results of
the calculations are shown at 5 Gyr intervals in Figure 3. Initially,
the particles (Panels (a) and (b)) have velocity dispersions and
density falloff corresponding exactly (within the statistical errors)
to the SLFC model. The evolution of the density and velocity
dispersions is shown over the 15 Gyr in panels (c)---(h). The velocity
dispersions changed quickly (in less than a Gyr), and the test
ensemble reaches a stationary phase-mixed state in which the
velocity distribution has become everywhere more isotropic relative to the
initial conditions.  The density falloff is very insensitive to these
changes, remaining virtually unchanged and well fitted by a $r^{-3.4}$
power-law (indicated by the lines in panels (d), (f) and (h)) during
the entire simulation.

  Motivated by this decrease of the velcoity anisotropy with time, we
performed the following experiment: we increased the initial
velocity anisotropy as a function of $r$, by replacing the SLFC
parameters $\sigma_o^2$ and $\sigma_+^2$ by the primed quantities
${\sigma^\prime}_o^2$ and ${\sigma^\prime}_+^2$. (These parameters
characterize the radial velocity variances in the inner and outer
halo model; see SLFC's eqs (3) and (4) and SLFC Figure 2).

$$ {\sigma^\prime}_+^2 = \lambda \sigma_+^2, ~~~~~~ \lambda \ge 1$$

\noindent and 

$$ {\sigma^\prime}_o^2 = \sigma_o^2 +
{1\over{2}}(1.0-\lambda)\sigma_+^2 ~~~~~~ \lambda \ge 1$$

  The effect of this transformation is best seen by looking at the
dotted line in Figure 4(a) (for the case $\lambda=1.6$) --- the
velocity anisotropy is increased in the inner and outer regions of the
halo, relative to the SLFC model.  We then performed the simulations
again, for different values of $\lambda$, corresponding to different
initial conditions.

\begin{figure}
\input epsf
\centering
\leavevmode
\epsfxsize=0.85
\columnwidth
\epsfbox{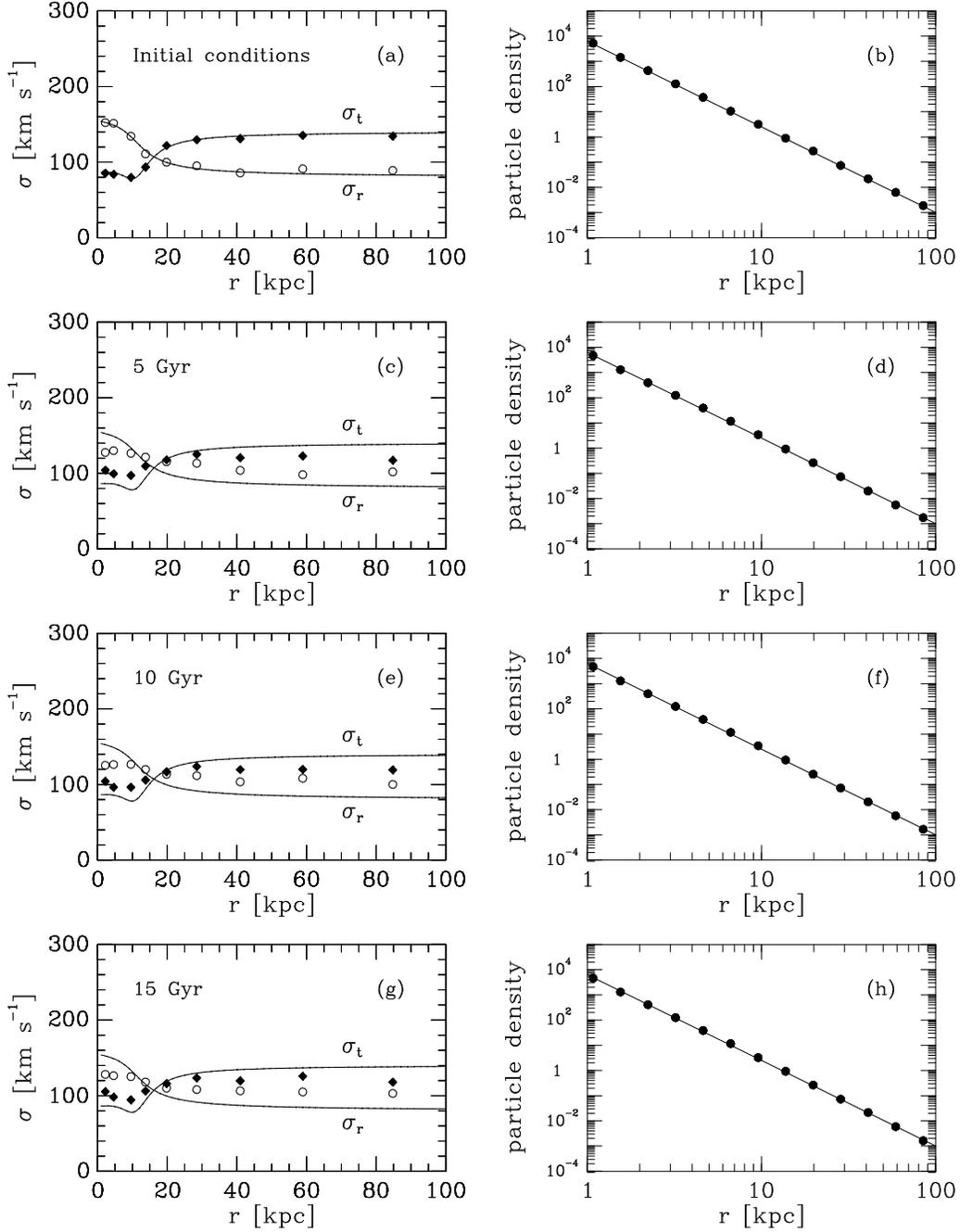}
\caption{The response of the particles to the
potential as a function of time is shown by their radial and
tangential velocity dispersions, $\sigma_r$ and $\sigma_t$ (lefthand
panels) and by their number density (kpc$^{-3}$) (righthand panels),
both as functions of Galactocentric radius $r$.  The particles
initially have $\sigma_r$ (open symbols) and $\sigma_t$ (solid
symbols) as specified by the SLFC model (indicated by solid lines),
and the random velocities of the particles are initially Gaussian. The
system relaxes quickly to a more isotropic state.}
\end{figure}

\begin{figure}
\input epsf
\centering
\leavevmode
\epsfxsize=0.85
\columnwidth
\epsfbox{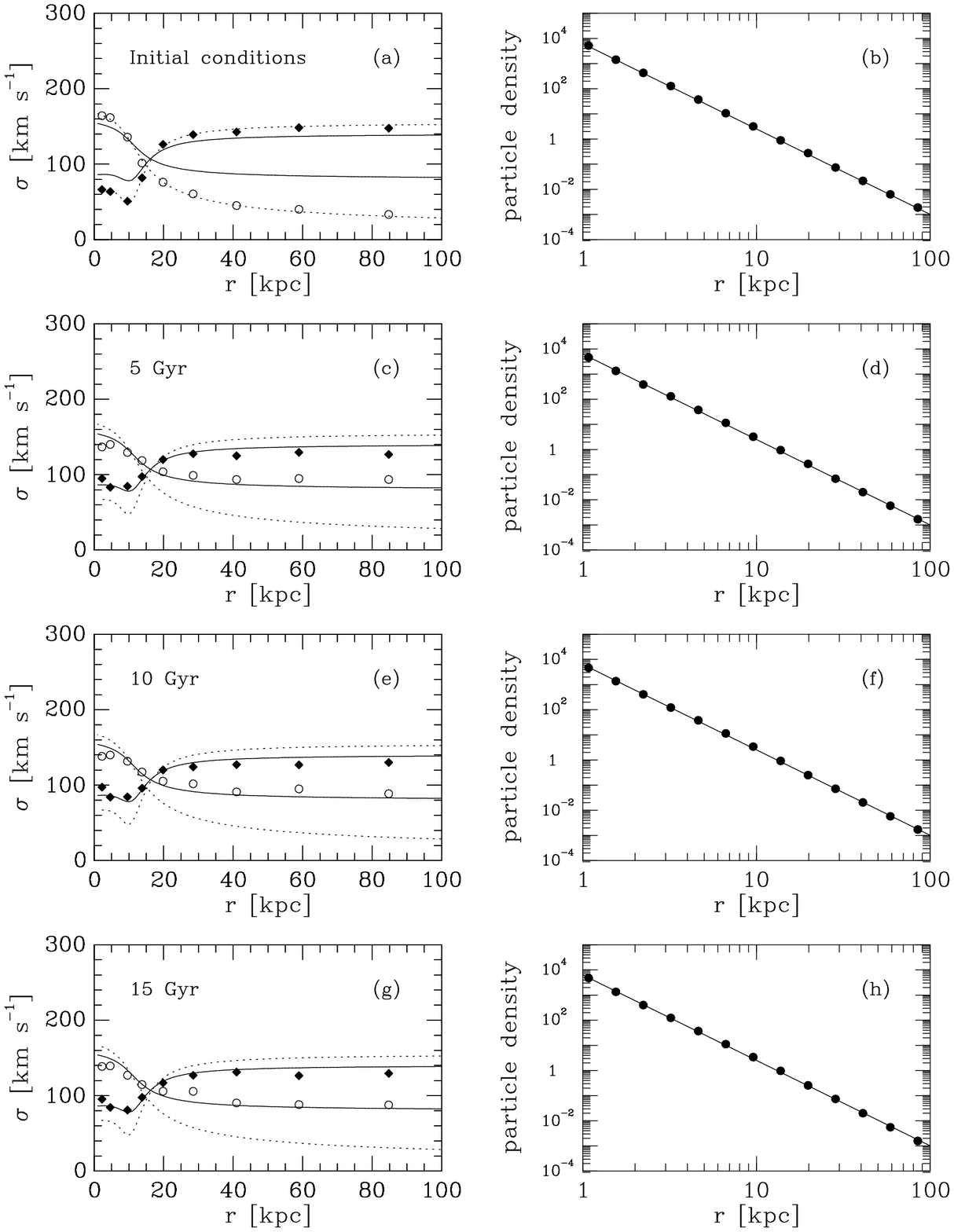}
\caption{Symbols are as for Figure 3. Here, the
initial anisotropy of the system has been artificially enhanced as is
described in section 3, and as indicated by the dashed lines. The
particles relax quickly and become a quite good stationary fit to the
SLFC model (shown by solid lines). This procedure strongly indicates
that a stationary system, having the characteristics of the SLFC
model, can exist.}
\end{figure}

  This simple strategy yielded quite satisfactory results. In
particular, we show in Figure 4 for the most satisfactory case,
$\lambda=1.6$, the evolution of $\sigma_r$, $\sigma_t$ and density
over 15 Gyr. As before, the velocities evolve quickly ($<$ 1 Gyr) to a
more isotropic state, becoming a good match to the SLFC model.  The
density falloff remains virtually unaffected as before, and remains a
good fit to the $r^{-3.4}$ power law.  

  Two further physical properties of the simulation were examined.  In
the SLFC model, we assumed for simplicity that firstly, the density
falloff of the halo is spherically symmetric, and secondly, that the
two azimuthal velocity dispersions (in radial coordinates,
$\sigma_\phi$ and $\sigma_\theta$) were equal (and denoted by
$\sigma_t$). In the simulation, the initial state was given these
properties, but since the integrations were fully three-dimensional,
these properties were free to evolve with time. In order to check the
self consistency of the stationary phase-mixed result of the
simulation with the SLFC model, we examined both these properties as
functions of time.

  Firstly, Figure 5 shows the density contours of the particles in the
$(x,z)$-plane, after they have attained their stationary
configuration.  Hartwick (1984), using samples of RR Lyrae stars in
many directions, has shown that while the outer halo (here taken as $r
\ga 10-20$ kpc) is close to being spherically symmetric, the inner
halo may be quite flattened, and in fact Larsen and Humphries (1994)
have recently argued that the inner halo may be flattened by as much
as 2 to 1. In this paper however, we are primarily concerned with the
behaviour of the outer halo, where the SLFC model is constrained by
observations.  Figure 5 shows that the outer halo (i.e $r \ga 10-20$
kpc) does remain spherically symmetric in the simulation, consistent
with the SLFC assumption, but that the inner halo (particularly inside
the solar circle) has been somewhat flattened in response to the
potential, though not as much as Larsen and Humphries propose. We do
stress however, that the SLFC model primarily meant to describe the
outer halo (and is constrained by observations in the outer halo
only), so the behaviour of the simulation in the inner halo must be
regarded with due caution.

\begin{figure}
\input epsf
\centering
\leavevmode
\epsfxsize=0.65
\columnwidth
\epsfbox{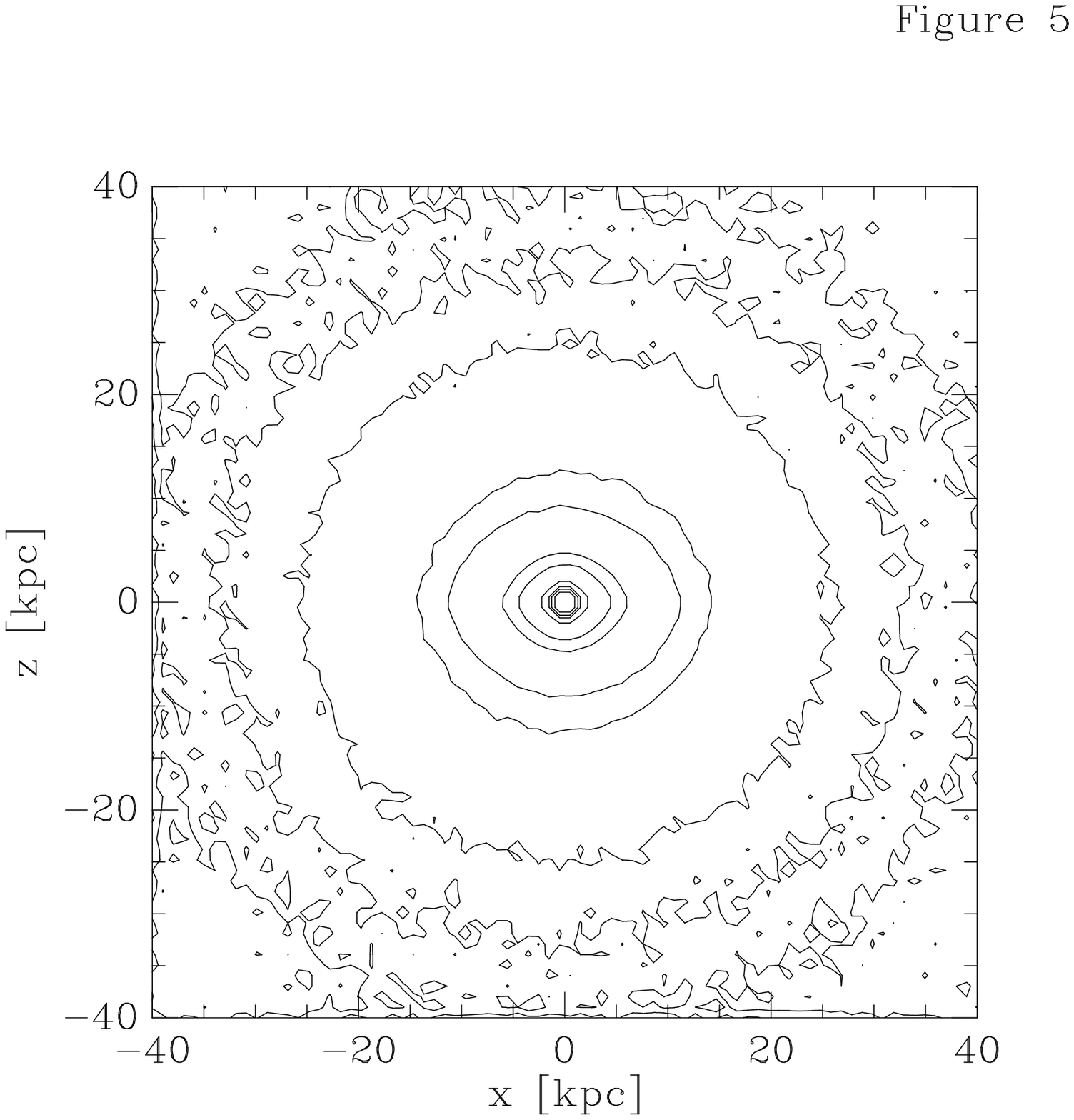}
\caption{The density distribution of the particles in
the simulation in Figure 4, shown in the $(x,z)$-plane, after the
model reaches its stationary state. The initial state of the
simulation was spherically symmetric. The outer parts of the
simulation are still spherical, while the inner part has flattened
somewhat in response to the potential.}
\end{figure}

\begin{figure}
\input epsf
\centering
\leavevmode
\epsfxsize=0.65
\columnwidth
\epsfbox{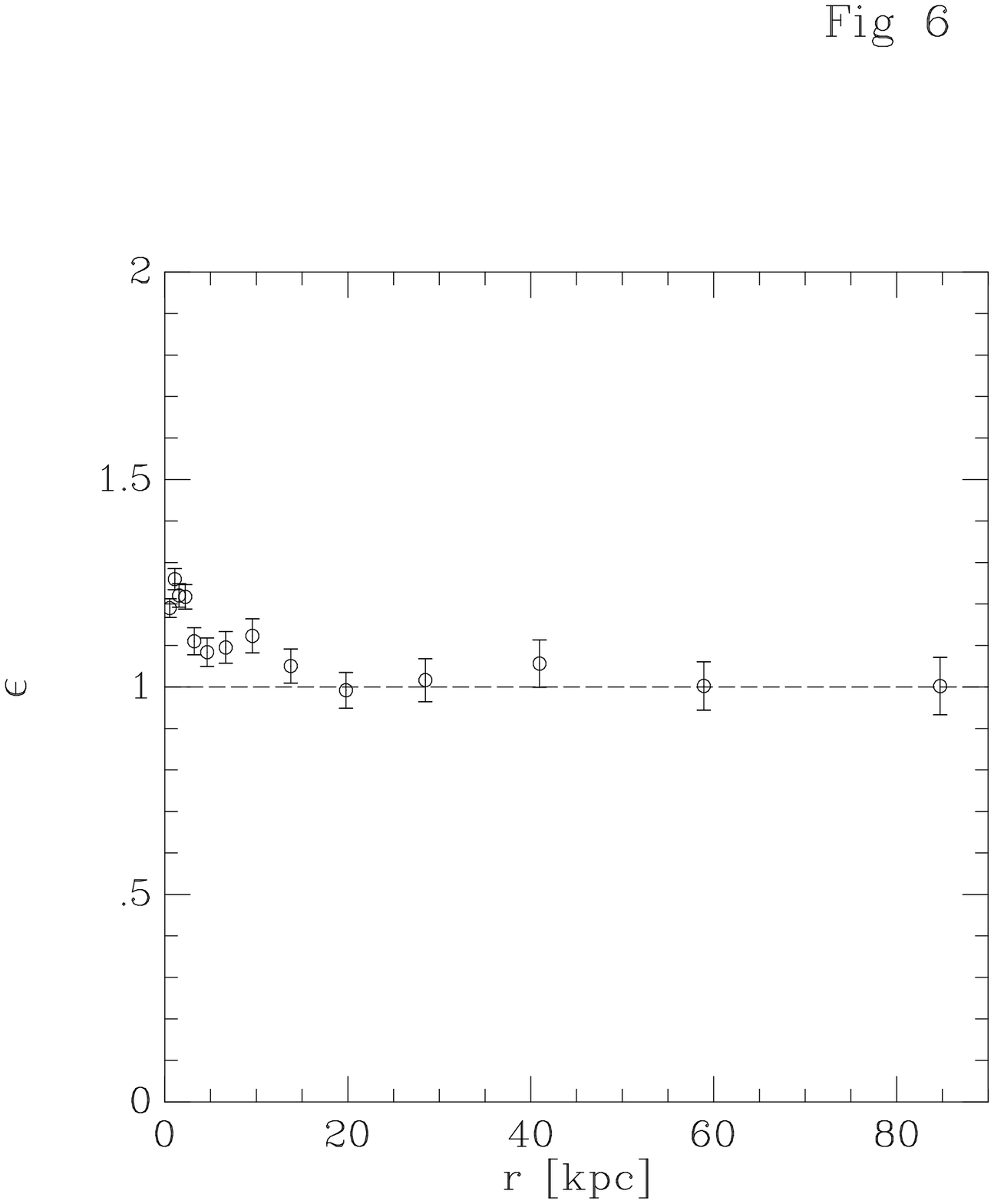}
\caption{The ratio $\epsilon=\sigma_\theta/\sigma_\phi$ as
a function of Galactocentric radius, $r$, after the simulation reaches
its stationary state. The SLFC model assumes that $\epsilon=1.0$ in the 
outer halo, and the simulation matches this well.}
\end{figure}

  Secondly, we examined the ratio $\epsilon=\sigma_\theta/\sigma_\phi$
which is shown as a function of Galactocentric radius $r$ in Figure 6,
again plotted after the model has attained a stationary state. By
assumption $\epsilon=1$ initially, but is free to evolve in the
simulation.  As with the density distribution, the inner and outer
regions responded differently. In the outer halo, $\epsilon$ remains a
good fit to the SLFC assumption of $\epsilon=1.0$, whereas $\epsilon$
departs slightly from this value in the inner halo. We stress again
that the SLFC data only constrain the behaviour of the outer halo, and
it is only in the outer halo that we require that $\epsilon$ should be
consistent with the SLFC model.

  In all the discussed simulations, the particles are kinematically
reorganized, arriving rapidly at a stationary state.  The response in
each case is primarily in the velocity distribution of the particles,
which evolves away from its initially Gaussian form, while the overall
density falloff with Galactocentric radius is hardly affected,
although some flattening occurs in the inner halo. If the particles
start with Gaussian velocity distribution with second moments exactly
matching the SLFC model, then the kinematic reorganization leads to a
stationary state which is no longer a match to the SLFC
model. However, for $\lambda=1.6$ the initially Gaussian velocity
distribution reorganizes itself into a new distribution which is a
close and stationary match to the SLFC model. This shows that at least
one velocity distribution exists with second moments matching the SLFC
model, and which remains stationary in the adopted potential.  Hence
we have a strong indication that the solutions of the SLFC model for
the velocity dispersions in the outer halo can be realized physically.

\section{Discussion and Conclusions}

  We have tested the simple model for the kinematics of the Galactic
halo (in particular the outer halo) proposed by SLFC, by directly
integrating particles in a realistic 3-D model of the Galactic
potential, under the assumption of an initially Gaussian velocity
distribution. We found that the particles relax somewhat into the
potential with time, the kinematics becoming everywhere more
isotropic.  We experimented with models where we increased the initial
anisotropy compared to the SLFC model, and found a configuration which
after a short relaxation period ($\approx$ 1 Gyr) becomes a quite good
and stationary fit to the SLFC model. Hence, the SLFC model, which
shows a notable change in the velocity anisotropy from markedly radial
at the sun to markedly tangential beyond about $r=20$ kpc, seems a
tenable description of outer halo kinematics.

  The origin of the change in the anisotropy with Galactocentric
radius remains unclear. We are currently gathering further data in the
outer halo at the poles and in the anti-center directions to test the
model more directly. We are also currently involved in a series of
hydrodynamical and N-body simulations of disk galaxy formation
including star-formation in which the kinematics of successive
generations of stars can be examined, and we expect the observed
variation in the velocity anisotropy with $r$ of the halo to be an
interesting constraint on the simulations.

\section*{Acknowledgments}

  We would like to thank Burkhard Fuchs for helpful discussions. CF is
very grateful to the Alexander von Humboldt Foundation (Bonn) for
funding and for the hospitality at the Astronomisches Rechen-Institut
in Heidelberg, where part of this research was carried out.  This work
was supported in part by Danmarks Grundforskningsfond through its
support of the Theoretical Astrophysics Center.

\section*{References}
\begin{trivlist}

\item[] Binney, J. and Tremaine, S. 1987. in {\it Galactic Dynamics},
Princeton University Press, Princeton, New Jersey.

\item[] Flynn, C. and Fuchs, B. 1994, MNRAS, 270, 471.

\item[] Flynn, C., Sommer-Larsen, J. and Christensen, P.R. 1994, MNRAS,
267, 77.

\item[] Flynn, C., Sommer-Larsen, J., Christensen, P.R. and Hawkins,
M.R.S., 1995, A\& ASupp, 109, 171.

\item[] Hartwick, F.D.A. 1987, in The Galaxy: (Reidel, Dordrecht)p. 281

\item[] Kuijken, K and Gilmore, G. 1989, MNRAS, 239, 651.

\item[] Larsen, J.A. and Humphries, R.M., 1994, ApJL, 436, L149.

\item[] Lewis, J. and Freeman, K.C., 1989, AJ, 97, 139.

\item[] Miyamoto, M. and Nagai, R., 1975, Pub. Ast. Soc. Jap. 27, 533.


\item[] Press, W.H., Flannery, B.P., Teukolsky, S.A. and
Vetterling, W.T. 1986, Numerical Recipes.

\item[] Sommer-Larsen, J., Flynn C. and Christensen, P.R. 1994, MNRAS,
271, 94.

\end{trivlist}

\end{document}